\documentclass[A4,10pt]{article} 

\usepackage{osameet3} 

\usepackage{subcaption}
\usepackage{caption}
\usepackage{pgfplots}
\usepackage{tikz}
\usetikzlibrary{positioning, shapes.geometric}
\usepackage{tikz,pgfplots,tkz-graph}%
\usepackage{amsmath}%
\pgfplotsset{compat=newest}%
\usetikzlibrary{positioning,matrix,shapes.multipart,shapes.misc,spy}%
\definecolor{lines-1}{RGB}{228,26,28}
\definecolor{lines-2}{RGB}{55,126,184}
\definecolor{lines-3}{RGB}{77,175,74}
\definecolor{lines-4}{RGB}{152,78,163}
\definecolor{lines-5}{RGB}{255,127,0}
\definecolor{lines-6}{RGB}{255,255,51}
\definecolor{lines-7}{RGB}{166,86,40}
\definecolor{lines-8}{RGB}{247,129,191}
\definecolor{lines-9}{RGB}{153,153,153}
\pgfplotscreateplotcyclelist{myCycleList}{
	color=lines-1,semithick,mark=o,mark size=1,mark repeat=10\\%
	color=lines-2,semithick,mark=star,mark size=1,mark repeat=10\\%
	color=lines-3,semithick,mark=square,mark size=1,mark repeat=10\\%
	color=lines-4,semithick,mark=diamond,mark size=1,mark repeat=10\\%
	color=lines-5,semithick,mark=triangle,mark size=1,mark repeat=10\\%
	color=lines-8,semithick,mark=Mercedes star,mark size=1,mark repeat=10\\%
	color=lines-7,semithick,mark=|,mark size=1,mark repeat=10\\%
	color=lines-9,semithick,mark=oplus star,mark size=1,mark repeat=10\\%
}
\pgfplotsset{
	compat=1.14,
	width =\columnwidth, 
	height=.8\columnwidth,
	ylabel absolute, ylabel style={yshift=-0.2cm},
	xlabel absolute, xlabel style={yshift=0.2cm},
	label style={font=\scriptsize},
	tick label style={font=\scriptsize},
	legend style={font=\scriptsize,cells={align=left}},
	grid=both,
	minor grid style={dotted},
}

\usetikzlibrary{spy}
\usepackage{amsmath,amssymb}
\usepackage[colorlinks=true,bookmarks=false,citecolor=blue,urlcolor=blue]{hyperref} 

\begin{document}

\title{Improved Simulation Accuracy \\
of the Split-Step Fourier Method}

\author{Shen Li\textsuperscript{1}, Magnus Karlsson\textsuperscript{2}, Erik Agrell\textsuperscript{1}}
\address{\textsuperscript{1}Dept. of Electrical Engineering, 
\textsuperscript{2}Dept. of Microtechnology and Nanoscience,\\ Chalmers University of Technology, SE-412 96 Gothenburg, Sweden}
\email{shenl@chalmers.se}
\copyrightyear{2020}
\begin{abstract}
We investigate a modified split-step Fourier method (SSFM) by including low-pass filters in the linear steps. This method can simultaneously achieve a higher simulation accuracy and a slightly reduced complexity.
\end{abstract}

\section{Introduction}
Signal propagation in fiber is generally governed by nonlinear Schr\"odinger equation (NLSE), which is a time-dependent nonlinear differential partial equation and cannot be solved analytically due to interactions between nonlinearity and linear dispersion \cite{WeidemanSSFM}. Thus, operational and accurate simulation algorithms are required to model the evolution of the electrical field in the fiber. The split-step Fourier method (SSFM) is the most commonly used way of simulating the NLSE because of its operability and high accuracy. In the SSFM, lightwave propagation along the fiber is discretized into many small spatial steps, in each of which the nonlinearity and dispersion can be separated and expressed analytically. Nevertheless, the SSFM usually requires high oversampling rates of the signal \cite{ErikCapacity,Kamran} and small step sizes to converge to the true result of the NLSE, indicating a trade-off between the simulation accuracy and complexity.

The computational complexity of the SSFM mostly comes from the large number of times of transformations between time domain and frequency domain using the fast Fourier transform and its inverse, and the exponential computations in the nonlinear operator. Much research has been devoted to analyzing the accuracy of different SSFM schemes and proposing refinements, particularly optimizing the selection and updating rules of the step size \cite{SimoneAccuracy,OlegVariable,QZhang,Heidt,Jshao}. In this paper, different from the all-pass filters usually used in the linear operators of the SSFM, we modify the SSFM by simply including a low-pass filter (LPF) in each linear operator to avoid possible aliasing during simulation. It shows that with these ``cost-free'' filters, the modeling accuracy of the SSFM is improved for a given set of oversampling rates and step sizes, or in other words, we can reduce the simulation complexity for a given accuracy. 

\section{Reduced-complexity SSFM}
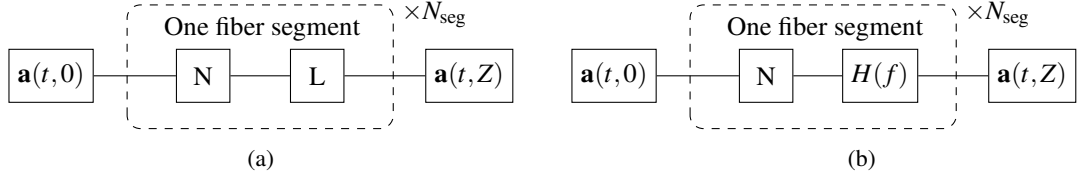
\begin{figure}
\centering
\label{Fig.1}

\begin{subfigure}[t]{0.48\textwidth}
\centering
\usetikzlibrary{shapes,arrows}
\tikzstyle{expr}=[font=\footnotesize]%
\tikzstyle{block} = [draw, fill=white, rectangle, 
    minimum height=2em, minimum width=2em]
\begin{tikzpicture}[auto, node distance=2cm,>=latex']
    \draw[black, dashed, rounded corners=5, fill=black!0] (1, -0.7) rectangle (4.5, 0.94);%
    \node [block] (a0) {$ \mathbf{a}(t,0) $};
    \node [block, right of=a0,node distance=2cm] (N) {N};
    \node [block, right of=N, node distance=1.5cm] (L) {L};
    \node [block, right of=L, node distance=2cm] (aZ) {$ \mathbf{a}(t,Z) $};
    \node[label=above:$\times N_\text{seg}$,above of=aZ, node distance=0.4cm, ] [xshift=-13pt] {};
    \node[label=above:{One fiber segment},above of=N, node distance=0.22cm, ] [xshift=22pt] {};
    \draw [-] (a0) -- node {} (N);
    \draw [-] (N) -- node {} (L);
    \draw [-] (L) -- node {} (aZ);
\end{tikzpicture}
\caption{}
\label{Fig.1sub.1}
\end{subfigure}
~
\begin{subfigure}[t]{0.48\textwidth}
\usetikzlibrary{shapes,arrows}
\tikzstyle{expr}=[font=\footnotesize]%
\tikzstyle{block} = [draw, fill=white, rectangle, 
    minimum height=2em, minimum width=2em]

\begin{tikzpicture}[auto, node distance=2cm,>=latex']
    \draw[black, dashed, rounded corners=5, fill=black!0] (1, -0.7) rectangle (4.5, 0.94);%
    \node [block] (a0) {$ \mathbf{a}(t,0) $};
    \node [block, right of=a0,node distance=2cm] (N) {N};
    \node [block, right of=N, node distance=1.5cm] (Hf) {$H(f)$};
    \node [block, right of=Hf, node distance=2cm] (aZ) {$ \mathbf{a}(t,Z) $};
    \node[label=above:$\times N_\text{seg}$,above of=aZ, node distance=0.4cm, ] [xshift=-13pt] {};
    \node[label=above:{One fiber segment},above of=N, node distance=0.22cm, ] [xshift=22pt] {};
    \draw [-] (a0) -- node {} (N);
    \draw [-] (N) -- node {} (Hf);
    \draw [-] (Hf) -- node {} (aZ);
\end{tikzpicture}
\caption{}
\label{Fig.1sub.2}
\end{subfigure}

\caption{Traditional SSFM structure (a) and improved SSFM with LPFs (b).}
\label{Fig.1}
\end{figure}

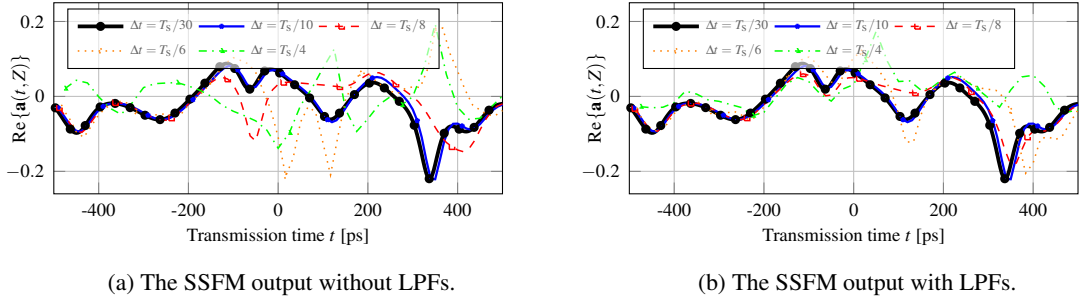
\begin{figure}
\centering
\begin{subfigure}[t]{0.47\textwidth}

\begin{tikzpicture}
	\begin{axis}[
		xmin=3.2, xmax=4.2,
		xticklabels={-400,-200,0,200,400},
		xtick={3.3,3.5,3.7,3.9,4.1},
		ymin=-0.26, ymax=0.25,
		xlabel={Transmission time $t$ [\text{ps}]},
		ylabel={$\text{Re}\left \{\mathbf{a}(t,Z)  \right \}$},
		ylabel style={at={(axis description cs:0.05,0.5)}, anchor=north},
		cycle list name=myCycleList,
		legend pos=north west,
		legend columns=3,
		legend entries = {$\Delta t=T_{\text{s}}/30$, $\Delta t=T_{\text{s}}/10$,$\Delta t=T_{\text{s}}/8$,$\Delta t=T_{\text{s}}/6$, $\Delta t=T_{\text{s}}/4$},
		legend style={fill=white, fill opacity=0.6, draw opacity=1,text opacity=1,font=\tiny},
		legend cell align=left,
		ylabel style={yshift=-.05cm},
		height =0.55\textwidth,
	]
	
	\addplot+[color=black, line width=0.5mm] table[
		x=t,
		y=y,
	] {./fig2data_30.txt};
	\addplot+[color=blue, line width=0.3mm] table[
		x=t,
		y=y,
	] {./fig2data_10.txt};
	\addplot+[dashed,color=red, line width=0.2mm] table[
		x=t,
		y=y,
	] {./fig2data_8.txt};
	\addplot+[dotted,color=orange, line width=0.2mm] table[
		x=t,
		y=y,
	] {./fig2data_6.txt};
	\addplot+[dash dot,color=green, line width=0.2mm] table[
		x=t,
		y=y,
	] {./fig2data_4.txt};
	\end{axis}

\end{tikzpicture}
\caption{The SSFM output without LPFs.}
\label{Fig.2sub.1}
\end{subfigure}
\begin{subfigure}[t]{0.47\textwidth}
\begin{tikzpicture}
	\begin{axis}[
		xmin=3.2, xmax=4.2,
		xticklabels={-400,-200,0,200,400},
		xtick={3.3,3.5,3.7,3.9,4.1},
		ymin=-0.26, ymax=0.25,
		xlabel={Transmission time $t$ [\text{ps}]},
		ylabel={$\text{Re}\left \{\mathbf{a}(t,Z)  \right \}$},
		ylabel style={at={(axis description cs:0.05,0.5)}, anchor=north},
		cycle list name=myCycleList,
		legend pos=north west,
		legend columns=3,
		legend entries = {$\Delta t=T_{\text{s}}/30$, $\Delta t=T_{\text{s}}/10$,$\Delta t=T_{\text{s}}/8$,$\Delta t=T_{\text{s}}/6$, $\Delta t=T_{\text{s}}/4$},
		legend style={fill=white, fill opacity=0.6, draw opacity=1,text opacity=1,font=\tiny},
		legend cell align=left,
		ylabel style={yshift=-.05cm},
		height =0.55\textwidth,
	]
	
	\addplot+[color=black, line width=0.5mm] table[
		x=t,
		y=y,
	] {./fig2data_30_filter.txt};
	\addplot+[color=blue, line width=0.3mm] table[
		x=t,
		y=y,
	] {./fig2data_10.txt};
	\addplot+[dashed,color=red, line width=0.2mm] table[
		x=t,
		y=y,
	] {./fig2data_8_filter.txt};
	\addplot+[dotted,color=orange, line width=0.2mm] table[
		x=t,
		y=y,
	] {./fig2data_6_filter.txt};
	\addplot+[dash dot,color=green, line width=0.2mm] table[
		x=t,
		y=y,
	] {./fig2data_4_filter.txt};
	\end{axis}

\end{tikzpicture}
\caption{The SSFM output with LPFs.}
\label{Fig.2sub.2}
\end{subfigure}

\caption{The output with different $\Delta t$ become more similar to the NLSE using LPFs. Parameters: $16$-QAM single wavelength transmission at $9.6$ dBm launch power; one span of $600$ km single mode fiber with ideal distributed amplification; $\beta_{2}=-21.7$ $\text{ps}^{2}$/km, $\gamma=1.27 (\text{W}\cdot \text{km})^{-1}$; step size $1.5$ km. The filter bandwidth is optimized for each $\Delta t$.}
\label{Fig.2}
\end{figure}

Let $\mathbf{a}(t,z)$ be the electrical field propagating along the fiber at time $t$ and distance $z$. The NLSE for the evolution of $\mathbf{a}(t,z)$ in an unamplified fiber span can be written as
\begin{eqnarray}
 \frac{\partial \mathbf{a}(t,z)}{\partial z}=-\frac{\alpha }{2}\mathbf{a}(t,z)-j\frac{\beta _{2}}{2}\frac{\partial ^{2}\mathbf{a}(t,z)}{\partial t^{2}}+j\gamma \left | \mathbf{a}(t,z) \right |^{2}\mathbf{a}(t,z)
\label{eq1} 
\end{eqnarray}
where $\mathbf{a}(t,0)$ is the input signal, $\alpha$ is the attenuation factor, $\beta_{2}$ is the dispersion parameter, and $\gamma$ is the nonlinear parameter. Let $L$ and $N$ denote the linear operator $-\frac{\alpha }{2}-j\frac{\beta _{2}}{2}\frac{\partial ^{2}}{\partial t^{2}}$ and nonlinear operator $j\gamma \left | \mathbf{a}(t,z) \right |^{2}$ in each step of the SSFM respectively. With $N_{\text{seg}}$ discretized steps and total transmission distance $Z$, Fig.~\ref{Fig.1sub.1} shows a traditional SSFM structure. The need for small enough time resolution $\Delta t$ can be illustrated by Fig.~\ref{Fig.2sub.1}, where $T_{\text{s}}$ is the symbol time. In this example, where the launch power is set high to stress-test the SSFM under adverse conditions, the SSFM output converges to the NLSE with $30$ samples per symbol and as $\Delta t$ increases, the output becomes increasingly deviated from the NLSE. After increasing $\Delta t$ to $T_{\text{s}}/4$, the output is completely independent of the NLSE. This phenomenon is caused by spectrum aliasing resulting from spectral broadening, since higher $\Delta t$ implies lower sampling frequency.

We modify the linear step in traditional SSFM by including an LPF with bandwidth $W$, as shown in Fig.~\ref{Fig.1sub.2}, to reduce aliasing. The proposed linear step is a multiplication with
\begin{equation}
H(f)=\left\{\begin{matrix}
 \exp(-\alpha \Delta z+2j\pi^{2}f^{2}\beta_{2}\Delta z),&  \left |f \right |\leq W\\ 
 0,& W< \left |f \right |\leq\frac{W_{\text{s}}}{2}
\end{matrix}\right.
\end{equation}
in the frequency domain, where $f$ is the frequency component of the signal, $\Delta z= Z/N_{seg}$ is the step size, $W_{\text{s}}$ is the sampling rate and the filter bandwidth $W\leq W_{\text{s}}/2$. When $W=W_{\text{s}}/2$, $H(f)$ implies the linear step in traditional SSFM. Thus, since fewer complex exponentials need to be computed when $W<W_{\text{s}}/2$, the proposed scheme has slightly lower complexity than the traditional one. The LPF bandwidth $W$ intuitively cannot be too narrow to avoid erasing too much information of the signal. Importantly, the purpose of the modified linear step is to improve the simulation accuracy for the standard NLSE channel \eqref{eq1}, which we do not modify in any way. This makes our contribution fundamentally different from, e.g., \cite{GarciaBrickwall}, where a new kind of propagation channel is created and analyzed by inserting band-pass filters regularly along the fiber. Our motivation is that the LPF improves the simulation accuracy, provided the filter bandwidth is carefully selected and optimized, which can be seen in Fig.~\ref{Fig.2sub.2}.

\section{Numerical results}
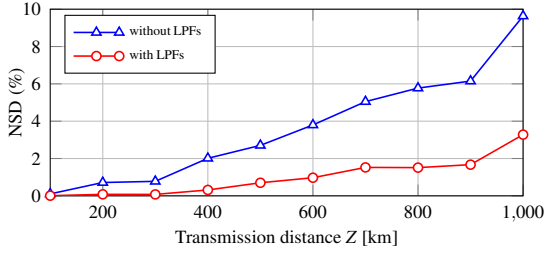
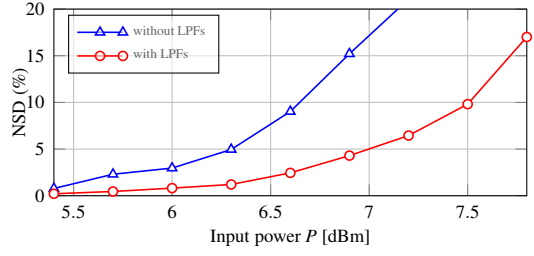
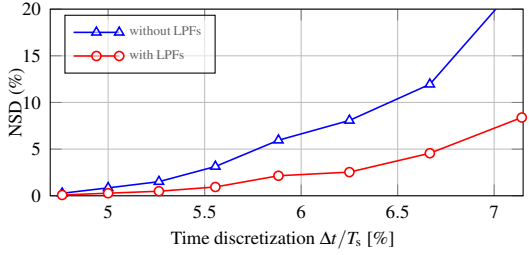
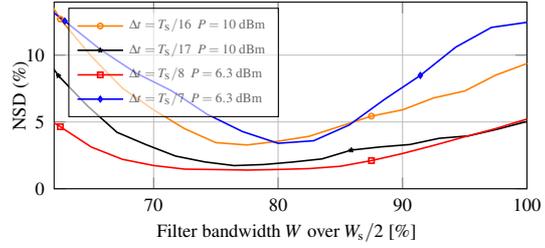
\begin{figure}
\centering
\begin{subfigure}[t]{0.49\textwidth}

\begin{tikzpicture}
	\begin{axis}[
		xmin=100, xmax=1000,
		ymin=0, ymax=10,
		xlabel={Transmission distance $Z$ [\text{km}]},
		ylabel={NSD (\%) },
		ylabel style={at={(axis description cs:0.05,0.5)}, anchor=north},
		cycle list name=myCycleList,
		legend pos=north west,
		legend entries = {without LPFs, with LPFs},
		legend style={font=\tiny},
		legend cell align=left,
		ylabel style={yshift=-.05cm},
		height =0.52\textwidth,
	]
	
	\addplot+[color=blue, mark=triangle*, mark options={mark size=2.0pt, fill=white, mark repeat=1}] table[
		x=Z,
		y=without,
	] {./fig3adata1km16.txt};
	\addplot+[color=red, mark=*, mark options={mark size=1.7pt, fill=white, mark repeat=1}] table[
		x=Z,
		y=with,
	] {./fig3adata1km16.txt};
	\end{axis}

\end{tikzpicture}
\caption{Parameters: $\Delta z =0.1$ km, $P=10$ dBm, $\Delta t = T_{\text{s}}/16$.}
\label{Fig.3sub.1}
\end{subfigure}
\begin{subfigure}[t]{0.49\textwidth}

\begin{tikzpicture}
	\begin{axis}[
		xmin=5.4, xmax=7.8,
		ymin=0, ymax=20,
		xlabel={Input power $P$ [\text{dBm}]},
		ylabel={NSD (\%) },
		ylabel style={at={(axis description cs:0.05,0.5)}, anchor=north},
		cycle list name=myCycleList,
		legend pos=north west,
		legend entries = {without LPFs, with LPFs},
		legend style={fill=white, fill opacity=0.6, draw opacity=1,text opacity=1,font=\tiny},
		legend cell align=left,
		ylabel style={yshift=-.05cm},
		height =0.52\textwidth,
	]
	
	\addplot+[color=blue, mark=triangle*, mark options={mark size=2.0pt, fill=white,mark repeat=1}] table[
		x=P,
		y=without,
	] {./fig3bdata4.txt};
	\addplot+[color=red, mark=*, mark options={mark size=1.7pt, fill=white,mark repeat=1}] table[
		x=P,
		y=with,
	] {./fig3bdata4.txt};
	\end{axis}
\end{tikzpicture}
\caption{Parameters: $\Delta z =0.1$ km, $\Delta t = T_{\text{s}}/8$, $Z=1000$ km.}
\label{Fig.3sub.2}
\end{subfigure}
\begin{subfigure}[t]{0.49\textwidth}
\begin{tikzpicture}
	\begin{axis}[
		xmin=4.7, xmax=7.15,
		ymin=0, ymax=20,
		xlabel={Time discretization $\Delta t/T_{\text{s}}$ [$\%$] },
		ylabel={NSD (\%) },
		ylabel style={at={(axis description cs:0.05,0.5)}, anchor=north},
		cycle list name=myCycleList,
		legend pos=north west,
		legend entries = {without LPFs, with LPFs},
		legend style={fill=white, fill opacity=0.6, draw opacity=1,text opacity=1,font=\tiny},
		legend cell align=left,
		ylabel style={yshift=-.05cm},
		xlabel style={xshift=-.05cm},
		height =0.52\textwidth,
	]
	
	\addplot+[color=blue, mark=triangle*, mark options={mark size=2.0pt, fill=white,mark repeat=1}] table[
		x=dt,
		y=without,
	] {./fig3cdata3.txt};
	\addplot+[color=red, mark=*, mark options={mark size=1.7pt, fill=white,mark repeat=1}] table[
		x=dt,
		y=with,
	] {./fig3cdata3.txt};
	
	\end{axis}
\end{tikzpicture}
\caption{Parameters: $\Delta z =0.1$ km, $P=10$ dBm, $Z = 1000$ km.}
\label{Fig.3sub.3}
\end{subfigure}
\begin{subfigure}[t]{0.49\textwidth}

\begin{tikzpicture}
	\begin{axis}[
		xmin=62, xmax=100,
        xticklabels={60,70,80,90,100},
		xtick={60,70,80,90,100},
		ymin=0, 
		xlabel={Filter bandwidth $W$ over $W_{\text{s}}/2$ [\text{\%}]},
		ylabel={NSD (\%) },
		ylabel style={at={(axis description cs:0.05,0.5)}, anchor=north},
		cycle list name=myCycleList,
	    legend pos=north west,
		legend entries = {$\Delta t=T_{\text{s}}/16$~~$P=10~\text{dBm}$, $\Delta t=T_{\text{s}}/17$~~$P=10~\text{dBm}$,$\Delta t=T_{\text{s}}/8$~~$P=6.3~\text{dBm}$,$\Delta t=T_{\text{s}}/7$~~$P=6.3~\text{dBm}$},
		legend cell align=left,
		legend style={fill=white, fill opacity=0.6, draw opacity=1,text opacity=1,font=\tiny},
		ylabel style={yshift=-.05cm},
		xlabel style={xshift=-.05cm},
		height =0.52\textwidth,
	]
	
	\addplot+[color=orange] table[
		x=BW,
		y=NSD,
	] {./fig3edata1km16P10.txt};
	\addplot+[color=black] table[
		x=BW,
		y=NSD,
	] {./fig3edata1km17.txt};
	\addplot+[color=red,] table[
		x=BW,
		y=NSD,
	] {./fig3edata1km8P6d3.txt};
	\addplot+[color=blue,] table[
		x=BW,
		y=NSD,
	] {./fig3edata1km76d3.txt};
	\end{axis}
\end{tikzpicture}
\caption{Parameters: $\Delta z=0.1$ km, $Z=1000$ km.}
\label{Fig.3sub.5}
\end{subfigure}

\caption{The NSD is improved with LPFs as a function of transmission distance $Z$, input power $P$, time discretization $\Delta t$, and filter bandwidth $W$. }
\label{Fig.3}
\end{figure}

We consider a $16$-QAM modulation format transmitted at $10$ Gbaud through single-mode fiber using a raised-cosine pulse with roll-off factor $10\%$ and ideal distributed amplification. The fiber parameters $\beta_{2}$ and $\gamma$ are $-21.7$ $\text{ps}^{2}$/km and $1.27 (\text{W}\cdot \text{km})^{-1}$, respectively, and the amplifier noise is neglected. To evaluate the accuracy, we define the normalized square difference (NSD)
\begin{equation}
\text{NSD}=\frac{\int (\mathbf{a}(t,Z)-\hat{\mathbf{a}}(t,Z))^{2}dt}{\int\mathbf{a}^{2}(t,Z)dt}
\end{equation}
between the NLSE output $\mathbf{a}(t,Z)$ and another simulated output $\hat{\mathbf{a}}(t,Z)$. Let $\Delta t_{\text{NLSE}}=T_{\text{s}}/30$ and $\Delta z_{\text{NLSE}}=0.1$ km denote the time discretization and step size of the benchmark NLSE. It was numerically validated that in all figures of this paper, these parameters are sufficient for the SSFM to converge to the NLSE output $\mathbf{a}(t,Z)$. We measure the NSD without and with LPFs as a function of: (a) transmission distance $Z$ [km], (b) input power $P$ [dBm], and (c) time discretization $\Delta t/T_{\text{s}}$ [\%] for some example cases as shown in Fig. \ref{Fig.3}. The NSD results in Fig. \ref{Fig.3} were obtained by averaging over many transmitted $16$-QAM symbol sequences. All the curves with the LPFs were obtained by globally searching the filter bandwidth $W$ to get the best performance. 

Fig. \ref{Fig.3sub.1} shows that at a given distance, the NSD reduces by a factor of $3$--$5$ using the LPFs, which means that with a fixed oversampling rate, we can simulate longer distances without losing accuracy. Similarly, Fig. \ref{Fig.3sub.2} implies that this modified SSFM shows more robustness to nonlinearity than the traditional SSFM. For a certain simulation time resolution, the NSD can be reduced by more than half using these filters (see Fig. \ref{Fig.3sub.3}). To this end, we can implement the SSFM with a larger sampling density with the LPFs and the only cost is to find the most suitable bandwidth to achieve the gain. The filter bandwidth $W$ over half of the sampling rate $W_{\text{s}}$ of the signal for some example cases is analyzed in Fig. \ref{Fig.3sub.5}, where the minimum NSD appears when $W$ is between $70\%$ and $85\%$, while $100 \%$ means the SSFM without LPFs. When $W$ is below $60\%$, the simulation becomes less accurate compared with the traditional SSFM, because the LPF erases some information of the signal.

\section{Conclusion and future work}
We proposed a modified SSFM algorithm with slightly lower complexity by including LPFs in the linear operator to avoid aliasing. How transmission distance, input power, and time discretization in the SSFM affect the NSD between the NLSE output and a simulated output is studied and the bandwidth of the LPFs is analyzed as well for some specific cases. We find that the proposed filter method can reduce more than $50\%$ of the simulation error for a wide range of link and simulation parameters.

For future work, it would be interesting if this filtering method could help with reducing the symbol error rate of the digital backpropagation algorithm, since the digital backpropagation is an inverse process of the SSFM where aliasing could happen as well.

\textbf{Acknowledgement}

This work was supported by the Swedish Research Council (VR) under grant no. 2017-03702.

\end{document}